\documentclass[useAMS,usenatbib]{mnras}

\usepackage[utf8]{inputenc}
\usepackage{graphicx}
\usepackage{color}
\usepackage{booktabs}
\usepackage[nolist]{acronym}
\usepackage{amssymb}
\usepackage[]{natbib} 
\usepackage{float}
\usepackage{hyperref}
\usepackage{amsmath,verbatim,mathtools,needspace,enumitem,etoolbox,physics,microtype}
\usepackage{epsfig}
\usepackage{url}
\usepackage{booktabs}
\usepackage{multirow}
\usepackage{array,multirow}

\newcolumntype{M}[1]{>{\centering\arraybackslash}m{#1}}
\newcolumntype{N}{@{}m{0pt}@{}}

\def\apj{ApJ}
\def\apjs{ApJS}
\def\apjl{ApJ}

\def\mnras{MNRAS}

\def\nat{Nature}
\def\araa{ARA\&A}
\def\aap{A\&A}

\def\pc{{\rm\thinspace pc}}
\def\kpc{{\rm\thinspace kpc}}

\def\Msun{\hbox{$\rm\thinspace M_{\odot}$}}

\def\yr{{\rm\thinspace yr}}

\def\Msunpc2{{\Msun\pc}^{-2}}
\def\Msunyrkpc2{{\Msun\yr^{-1}\kpc}^{-2}}

\def\magarcsec2{{\rm\thinspace mag\thinspace arcsec}^{-2}}

\definecolor{orange}{RGB}{248, 100, 0}

\definecolor{mypink2}{RGB}{219, 48, 122}

\definecolor{OOR}{RGB}{190, 103, 39}

\begin{document}

\title[New insights on BBH formation]{New insights on binary black hole formation channels after GWTC-2: young star clusters versus isolated binaries}

\author[Bouffanais et al.]{Yann Bouffanais$^{1,2}$\thanks{E-mail:bouffanais@pd.infn.it, yann.bouffanais@gmail.com},  Michela Mapelli$^{1,2,3}$\thanks{E-mail:michela.mapelli@unipd.it}, Filippo Santoliquido$^{1,2}$, Nicola Giacobbo$^{1,2,3,6}$, \newauthor Ugo N. Di Carlo$^{1,2,3,4}$ , Sara Rastello$^{1,2}$,  M. Celeste Artale$^{5}$, Giuliano Iorio$^{1,2}$ \\ 
$^{1}$Physics and Astronomy Department Galileo Galilei, University of Padova, Vicolo dell'Osservatorio 3, I--35122, Padova, Italy\\
$^{2}$INFN-Padova, Via Marzolo 8, I--35131 Padova, Italy\\
$^{3}$INAF-Osservatorio Astronomico di Padova, Vicolo dell'Osservatorio 5, I--35122, Padova, Italy\\
$^{4}$Dipartimento di Scienza e Alta Tecnologia, University of Insubria, Via Valleggio 11, I--22100, Como, Italy\\
$^{5}$Institut f{\"u}r  Astro- und Teilchenphysik, Universit{\"a}t Innsbruck, Technikerstrasse 25/8, A-6020, Innsbruck, {\"O}sterreich\\
$^{6}$School of Physics and Astronomy \& Institute for Gravitational Wave Astronomy, University of Birmingham, Birmingham, B15 2TT, UK}

\maketitle
\begin{abstract}

With the recent release of the second gravitational-wave transient catalogue (GWTC-2), which introduced dozens of new detections, we are at a turning point of gravitational wave astronomy, as we are now able to directly infer constraints on the astrophysical population of compact objects. Here, we tackle the burning issue of  understanding the origin of binary black hole (BBH) mergers. To this effect, we make use of state-of-the-art population synthesis and N-body simulations, to represent two distinct formation channels: BBHs formed in the field (isolated channel) and in young star clusters (dynamical channel). We then use a Bayesian hierarchical approach to infer the distribution of the mixing fraction $f$, with $f=0$ ($f=1$) in the pure dynamical (isolated) channel. 
We explore the effects of additional hyper-parameters of the model, such as the spread in metallicity $\sigma_{\text{Z}}$ and the  parameter $\sigma_{\text{sp}}$, describing the distribution of spin magnitudes.  We find that the dynamical model is slightly favoured with a median value of $f=0.26$, when   $\sigma_{\text{sp}}=0.1$ and $\sigma_{\text{Z}}=0.4$. Models with higher spin magnitudes tend to strongly favour dynamically formed BBHs ($f\le{}0.1$ if $\sigma_{\text{sp}}=0.3$). Furthermore, we show that hyper-parameters controlling the rates of the model, such as $\sigma_{\rm Z}$, have a large impact on the inference of the mixing fraction, which rises from  $0.18$ to $0.43$ when we increase $\sigma_{\text{Z}}$ from 0.2  to 0.6, for a fixed value of  $\sigma_{\text{sp}}=0.1$. Finally, our current set of observations is better described by a combination of both formation channels, as a pure dynamical scenario is excluded at the $99\%$ credible interval, except when the spin magnitude is high.

\end{abstract}


\section{Introduction}

In 2015, the LIGO--Virgo collaboration (LVC) announced the first direct detection of gravitational waves (GWs) emitted by the merger of a binary black hole (BBH, \citealt{abbottGW150914,abbottGW150914astro}). Recently, the LVC has published  a total of 50 binary compact object mergers, mostly BBHs, as a result of the first (O1, \citealt{abbottO1}), the second (O2,  \citealt{abbottO2,abbottO2pop}) and the first half of the third observing run (O3a, \citealt{abbottO3a,abbottO3apop}). Other authors \citep{zackay2019,udall2019,venumadhav2020,nitz2020,nitz2021} claim several additional BBH candidates, based on an independent analysis of the  LVC data.


This extraordinary wealth of data gives an unprecedented opportunity to understand the formation channels of BBHs (see, e.g., \citealt{mandel2018a} and \citealt{mapelli2018c} for a review). In the isolated formation channel, BBHs originate from the evolution of massive binary stars. A BBH that forms in isolation can reach coalescence within a Hubble time only if its stellar progenitors evolve via common envelope \citep[e.g., ][]{tutukov1973,bethe1998,portegieszwart1998,belczynski2002,belczynski2008,belczynski2016,eldridge2016,stevenson2017,mapelli2017,mapelli2019,kruckow2018,spera2019,tanikawa2020,belczynski2020}, stable mass transfer \citep[e.g.,][]{giacobbo2018,neijssel2019,bavera2020,bouffanais2020,andrews2020}, or chemically homogeneous evolution \citep[e.g.,][]{marchant2016,mandel2016,demink2016,dubuisson2020}. These processes tend to pose a limit of $\sim{90-100}$ M$_\odot$ to the total maximum mass of a BBH merger \citep[e.g.,][]{bouffanais2019} and to align the spins of the final black holes (BHs) with the orbital angular momentum of the binary system \citep[e.g.,][]{mandel2016,rodriguez2016spin,gerosa2018}. Only supernova explosions can partially misalign the systems \citep[e.g.,][]{kalogera2000}. 

Alternatively, BBHs can form dynamically: BHs can pair up with other  BHs in dense stellar systems, such as nuclear star clusters \citep[e.g.,][]{antonini2016,petrovich2017,antonini2019,Sedda_2020,fragione2020b}, globular clusters \citep[e.g.,][]{portegieszwart2000,tanikawa2013,rodriguez2016,askar2017,fragionekocsis2018,choksi2018,choksi2019,hong2018,rodriguezloeb2018}, and young star clusters \citep[e.g.,][]{banerjee2010,ziosi2014,mapelli2016,banerjee2017,banerjee2020,dicarlo2019a,dicarlo2020,kumamoto2019,kumamoto2020}. Finally, gas torques in AGN discs \citep[e.g.,][]{2017ApJ...835..165B,2017MNRAS.464..946S,mckernan2018,2019ApJ...876..122Y,tagawa2020} and hierarchical triples \citep[e.g.,][]{antonini2017,silsbee2017,fragione2020,vigna2021} can also facilitate  the formation of BBH mergers.
Unlike isolated BBHs, dynamical BBHs can have total masses in excess of $\sim{}100$ M$_\odot$ as a result of (runaway) stellar mergers in young star clusters \citep[e.g.,][]{portegieszwart2004,mapelli2016,dicarlo2019b,rizzuto2021} or hierarchical BBH mergers \citep[e.g.,][]{miller2002,giersz2015,fishbach2017,gerosa2017,rodriguez2019,mapelli2021}. Moreover, dynamics reset the memory of spin orientation: we expect that dynamical BBHs have isotropically oriented spins \citep[e.g.,][]{rodriguez2016spin}.

Based on these two main differences on the mass and spin distribution of dynamical versus isolated BBHs, several studies have explored the possibility of using LVC data to constrain the formation channels of BBHs \citep[e.g.,][]{mandel2016,zevin2017,stevenson2017,bouffanais2019,callister2020,zevin2020,wong2020,roulet2021}. Considering O1, O2 and O3a events, the LVC collaboration recently showed that 12--44\% of BBHs have spins tilted by more than 90$^\circ$ with respect to their orbital angular momentum (\citealt{abbottO3apop}). This suggests that isolated BBHs can hardly account for the entire sample. Here, we compare the properties of BBHs in the second GW transient catalogue (hereafter, GWTC-2, \citealt{abbottO3a,abbottO3apop}) against our population synthesis and dynamical simulations. We consider isolated binary evolution and dynamical formation in young dense star clusters \citep{santoliquido2020}. We show that GWTC-2 data strongly disfavour the possibility that all observed BBHs come from isolated binary evolution, even when we take into account the main uncertainties on metallicity evolution and spin magnitudes.

This paper is organised as follows. Section~\ref{sec:methods} describes our astrophysical models. In Section~\ref{sec:Bayes}, we summarize our Bayesian hierarchical analysis. We present and discuss our results in Sections~\ref{sec:results} and \ref{sec:discussion}, respectively. A short summary is provided in Section~\ref{sec:summary}.

\section{Astrophysical model}\label{sec:methods}


\subsection{Isolated BBHs}
\label{sec_MOBSE} 

The isolated BBHs were simulated with the population synthesis code\footnote{\url{http://demoblack.com/catalog_codes/mobse-public-version/}} {\sc mobse} \citep{giacobbo2018,mapelli2018}. With respect to its progenitor {\sc bse} \citep{hurley2000,hurley2002}, {\sc mobse} contains updated prescriptions for stellar winds. In particular, mass loss by massive hot stars is expressed as $\dot{M}\propto{}Z^\gamma{}$, where $Z$ is the stellar metallicity and $\gamma{}$ is a function of the stellar luminosity through the Eddington ratio  \citep{giacobbo2018b}. In {\sc mobse}, the mass of a compact object depends on the final total mass and on the final carbon-oxygen core mass of its progenitor star, as described in \cite{fryer2012}. Here, we adopt the rapid core-collapse supernova model from \cite{fryer2012}, which enforces a compact-object mass gap between 2 and 5 M$_\odot$. Electron-capture supernovae are implemented as described in \cite{giacobbo2019}. Pulsational pair instability and pair-instability supernovae are modelled as in \cite{mapelli2020}. These models yield a maximum BH mass of $\approx{}65$ M$_\odot$. 


We assign natal kicks to the BHs as $v_{\rm BH}=v_{\rm Max}\,{}(1-f_{\rm fb})$, where  $f_{\rm fb}$ is the fraction of fallback as defined in \cite{fryer2012}, while $v_{\rm Max}$ is a random number drawn from a Maxwellian distribution with one-dimensional root-mean square velocity $\sigma{}$ \citep{hobbs2005}. In this manuscript, we adopt $\sigma{}=15$ km s$^{-1}$. 
The main binary evolution processes (mass transfer, common envelope and tides) are described as in \cite{hurley2002}. For common envelope, we adopt the $\alpha{}$ formalism with $\alpha{} = 5$, while the parameter $\lambda{}$ is calculated self-consistently with the prescriptions derived by \cite{claeys2014}. Orbital decay by GW emission is implemented as in \cite{peters1964}. All BBHs evolve by GW emission, regardless of their orbital separation.

BH spin magnitudes are drawn from a Maxwellian distribution with $\sigma_{\rm sp}=0.1$, in the fiducial case. We also discuss the two extreme cases with $\sigma_{\rm sp}=0.01$ and 0.3. Theoretical models of BH spin magnitudes are affected by substantial uncertainties, mostly because of our poor understanding of angular momentum transfer in the stellar interior \citep[e.g.,][]{belczynski2020}. The case with $\sigma_{\rm sp}=0.01$ corresponds to assuming extremely efficient angular momentum dissipation \citep{fullerma2019}, while values of $\sigma{}_{\rm sp}=0.1-0.3$ are more conservative.

In isolated binaries, the spins tend to be re-aligned by mass transfer and tides (but see \citealt{stegmann2020} for a possible spin flip mechanism during mass transfer). Supernovae can produce a tilt between the final and the initial orbital angular momentum direction, hence they can induce a misalignment between the BH spins and the orbital angular momentum vector. In our simulations, we calculate this tilt as 
\begin{equation}\label{eq:spin}
\cos{\theta{}}=\cos{(\nu{}_1)}\,{}\cos{(\nu{}_2)}+\sin{(\nu{}_1)}\,{}\sin{(\nu{}_2)}\,{}\cos{(\phi{})},
\end{equation}
where $\nu{}$ is the angle between the orbital angular momentum vector after ($\vec{L}_{\rm new}$) and before a supernova explosion ($\vec{L}_{\rm old}$), so that 
\begin{equation}
    \cos{(\nu{})}=\frac{\vec{L}_{\rm new}}{L_{\rm new}}\cdot{}\frac{\vec{L}_{\rm old}}{L_{\rm old}}. 
\end{equation}
In eq.~\eqref{eq:spin}, $\nu_1$ ($\nu_2$) corresponds to the tilt induced by the first (second) supernova, while $\phi$ is the phase of the projection of $\vec{L}$ into the orbital plane. Our formalism neglects possible re-alignments of the spin of the first born BH before the formation of the second BH. 
We get $\nu{}_1$ and $\nu{}_2$ directly from {\sc mobse}, while we have to generate $\phi$ as a uniform random number between 0 and $2\,{}\pi$ \citep{gerosa2013,rodriguez2016spin}.



We have simulated binary stars with 12 different metallicities: $Z = 0.0002$, 0.0004, 0.0008, 0.0012, 0.0016, 0.002, 0.004, 0.006, 0.008, 0.012, 0.016, 0.02. We have simulated $10^7$ binaries per each metallicity comprised between $Z = 0.0002$ and 0.002, and $2\times{}10^7$ binaries per each metallicity $Z\ge{}0.004$, since higher metallicities are associated with lower BBH merger efficiency (e.g. \citealt{giacobbo2018b,klencki2018}). Thus, we have simulated $1.8\times{}10^8$ isolated binaries. The zero-age main-sequence masses of the primary component of each binary star are distributed according to a Kroupa \citep{kroupa2001} initial mass function in the range $[5,\,{}150]\,{}{\rm M}_\odot$. The orbital periods, eccentricities and mass ratios of binaries are drawn from \cite{sana2012}. In particular, we derive the mass ratio $q$ as $\mathcal{F}(q) \propto q^{-0.1}$ with $q\in [0.1,\,{}1]$, the orbital period $P$ from $\mathcal{F}(\Pi) \propto \Pi^{-0.55}$ with $\Pi = \log{(P/\text{day})} \in [0.15,\,{} 5.5]$ and the eccentricity $e$ from $\mathcal{F}(e) \propto e^{-0.42}~~\text{with}~~ 0\leq e \leq 0.9$.

\subsection{Dynamical BBHs}
\label{sec_dyn}

The dynamical simulations have already been presented in previous work \citep{dicarlo2020,rastello2020,santoliquido2020}. Here, we summarize their main features, while we refer to the aforementioned papers for more details. We simulated dense young star clusters with total masses $M_{\rm SC}$ between 300 and $3\times{}10^4$ M$_\odot$, randomly drawn from a distribution ${\rm d}N/{\rm d}M_{\rm SC}\propto{}M^{-2}_{\rm SC}$, consistent with observations \citep{lada2003,portegieszwart2010}. In particular, we analyze $100002$ simulations of star clusters with mass $300-1000$ M$_\odot$ from \cite{rastello2020} and 6000 simulations of star clusters with mass $1000-30000$  M$_\odot$, corresponding to the union of set~A and~B presented in \cite{dicarlo2020}. The dynamical simulations are equally divided between three metallicities: $Z=0.02,$ 0.002 and 0.0002.

All the dynamical simulations were performed with the direct N-body code {\sc nbody6++gpu} \citep{wang2015,wang2016}, interfaced with our population synthesis code {\sc mobse} \citep{giacobbo2018}. This guarantees that the treatment of stellar and binary evolution is the same for isolated and dynamical BBHs. 


The global initial binary fraction of the dynamical simulations is $f_{\rm bin}=0.4$, but the binary fraction is assumed to correlate with mass \citep{kuepper2011}. As a result, all stars with masses $>5$ M$_\odot$ are initially members of a binary system \citep{dicarlo2019a}, consistent with observations \citep{sana2012,moe2017}. The orbital periods, eccentricities and mass ratios of the original binaries are drawn from the same distribution as the isolated binaries.

Spin magnitudes of dynamical BBHs are drawn from a Maxwellian distribution in the same way as we did for isolated BBHs. In particular, we adopt $\sigma_{\rm sp}=0.1$ as our fiducial case, and we also explore $\sigma_{\rm sp}=0.01$ and 0.3. Dynamical processes such as exchanges, flybys and captures tend to misalign the spins. Hence, we assume that all components of the dynamically formed BBHs have isotropic spins over the sphere.


\subsection{Redshift evolution and merger rate density}
\label{sec_merger_rate}

We calculated the merger rate density as
\begin{eqnarray}
\label{eq:mrd}
   \mathcal{R}(z) = \frac{{\rm d}~~~~~}{{\rm d}t_{\rm lb}(z)}\int_{z_{\rm max}}^{z}\left[\int_{Z_{\rm min}}^{Z_{\rm max}}\eta(Z) \,{}\mathcal{F}(z',z, Z)\,{}{\rm d}Z\right]\\\nonumber{}
   \times{}\psi(z')\,{}\frac{{\rm d}t_{\rm lb}(z')}{{\rm d}z'}\,{}{\rm d}z',
\end{eqnarray}
where $t_{\rm lb}(z)$ is the look-back time at redshift $z$, $Z_{\rm min}$ and $Z_{\rm max}$ are the minimum and maximum metallicity, $\psi{}(z')$ is the cosmic star formation rate (SFR) density at redshift $z'$. In eq.~\eqref{eq:mrd}, we used the fit to the SFR density from \cite{madau2017}:
\begin{equation}
\label{eq:sfrd}
 \psi(z) = 0.01\,{}\frac{(1+z)^{2.6}}{1+[(1+z)/3.2]^{6.2}}~\text{M}_\odot\,{}\text{Mpc}^{-3}\,{}\text{yr}^{-1}.
\end{equation}

In eq.~\eqref{eq:mrd}, $\eta(Z)$ is the merger efficiency, namely the ratio between the total number $\mathcal{N}_{\text{TOT}}(Z)$ of compact binaries (formed from a coeval population) that merge within an Hubble time ($t_{{\rm H}_0} \lesssim 14$ Gyr) and the total initial mass $M_\ast{}(Z)$ of the simulation with metallicity $Z$.

Finally, $\mathcal{F}(z',z,Z)$ is the fraction  of compact binaries that form at redshift $z'$ from stars with metallicity $Z$ and merge at redshift $z$:
\begin{equation}\label{eq:Fz}
\mathcal{F}(z',z,Z)=\frac{\mathcal{N}(z',z,Z)}{\mathcal{N}_{\text{TOT}}(Z)}\,{}p(z', Z),
\end{equation}
where $p(z', Z)$ is the distribution of stellar metallicities at redshift $z'$ and $\mathcal{N}(z',z,Z)$ is the total number of compact binaries that merge at redshift $z$ and form from stars with metallicity $Z$ at redshift $z'$.

The merger rate density is dramatically affected by the metallicity evolution of stars across cosmic time, which is needed to calculate the term $p(z',Z)$ of eq.~\eqref{eq:Fz}. Here we use the fit to the mass-weighted metallicity evolution given by \cite{madau2017}:
\begin{equation}\label{eq:met}
    \log{\langle{}Z/{\rm Z}_\odot\rangle{}}=0.153-0.074\,{}z^{1.34}
\end{equation}
To describe the spread around the mass-weighted metallicity, we assume that metallicities are distributed according to a log-normal distribution: 
\begin{equation}
\label{eq:pdf}
p(z', Z) = \frac{1}{\sqrt{2 \pi\,{}\sigma_{\rm Z}^2}}\,{} \exp\left\{{-\,{} \frac{\left[\log{(Z(z')/{\rm Z}_\odot)} - \log{\langle{}Z(z')/Z_\odot\rangle{}}\right]^2}{2\,{}\sigma_{\rm Z}^2}}\right\}.
\end{equation}

The standard deviation $\sigma{}_Z$ is highly uncertain. Here, we probe different values of $\sigma{}_Z=0.2,$ 0.4 and 0.6. In eq.~\eqref{eq:pdf}, we use $\log{\langle{}Z(z')/Z_\odot\rangle{}}$ instead of $\langle{}\log{(Z(z')/Z_\odot)}\rangle{}$ on purpose, because we want to show what happens if we let the mean value of the log-normal distribution unchanged and we just vary the value of the standard deviation $\sigma_{\rm Z}$. This is a reasonable simplification, given the large uncertainty on the average metallicity evolution and its spread. In Appendix~\ref{sec:appendix}, we show what we get if we use $\langle{}\log{(Z(z')/Z_\odot)}\rangle{}$ instead.

To calculate the merger rate density as a function of redshift in the comoving frame, we made use of the software {\sc cosmo}$\mathcal{R}${\sc ate}. We refer to \cite{santoliquido2020} and \cite{santoliquido2020b} for further details. 





\subsection{Analytical model description}

Our astrophysical models can be described as a function of a set of hyper-parameters $\lambda$. 
In this study, the hyper-parameters $\lambda$ are a combination of the formation channel type (either isolated or dynamical), the metallicity dispersion ($\sigma_{\rm Z}=\lbrace 0.2,\,{}0.4,\,{}0.6 \rbrace$) and the spin magnitude root-mean square ($\sigma_{\text{sp}} = \lbrace 0.01,\,{}0.1,\,{}0.3 \rbrace$).

A given model will have a prediction on the distribution of merging BBHs,
\begin{equation}
\dfrac{\text{d}N}{\text{d} \theta} (\lambda) = N_{\lambda}\,{} p(\theta | \lambda),
\label{eq_pop_BBH}
\end{equation}
where $\theta$ are the parameters of BBH mergers and $N_{\lambda}$ is the total number of mergers predicted by the model computed as
\begin{equation}
N_{\lambda} = \int_{0}^{z_{\text{h}}} \mathcal{R}(z)\,{} \dfrac{\text{d} V_{c}}{\text{d}z}\,{} \dfrac{T_{\text{obs}}}{1+z}\,{} \text{d}z,
\label{num_det}
\end{equation}
where $\text{d} V_{c} / \text{d}z$ is the comoving volume element and $T_{\rm obs}$ is the observation time considered in the analysis. The integral in eq.~\eqref{num_det} is done over redshift ranging from 0 to $z_{\rm h}$, where $z_{\rm h}$ is the horizon redshift, i.e. the redshift corresponding to the instrumental horizon of LIGO and Virgo. For this specific analysis, we selected a value $z_{\rm h} = 2$, as it sets a safe boundary for the detection limit of the detectors during the first three observing runs.

To model the population of merging BBHs, we have chosen the parameterisation $\theta = \lbrace \mathcal{M}_{\rm c},\,{}q,\,{}z,\,{}\chi_{\text{eff}} \rbrace$ where $\mathcal{M}_{\rm c}$ is the chirp mass and $\chi_{\text{eff}}$ the effective spin,
\begin{equation}
\chi_{\rm eff}= \frac{(m_1\,{}\vec{\chi}_1+m_2\,{}\vec{\chi}_2)}{m_1+m_2}\cdot{}\frac{\vec{L}}{L},
\end{equation}
where $m_1$ ($m_2$) is the primary (secondary) BH mass, $\chi_1$ ($\chi_2$) is the primary (secondary) dimensionless spin magnitude and $\vec{L}$ is the orbital angular momentum of the BBH.

To compute the distribution $p(\theta | \lambda)$, we first constructed catalogues of $3 \times 10^{5}$ sources for all possible combinations of hyper-parameters $\lambda$, using the merger rate and metallicity distribution calculated by {\sc cosmo$\mathcal{R}$ate}. From these catalogues, we can derive a continuous estimation of $p(\theta | \lambda)$ by making use of Gaussian kernel density estimation. A value of the bandwidth of $0.075$ proved to be our optimal choice to describe our distributions.

\begin{figure*}
\includegraphics[width = 0.9\textwidth ]{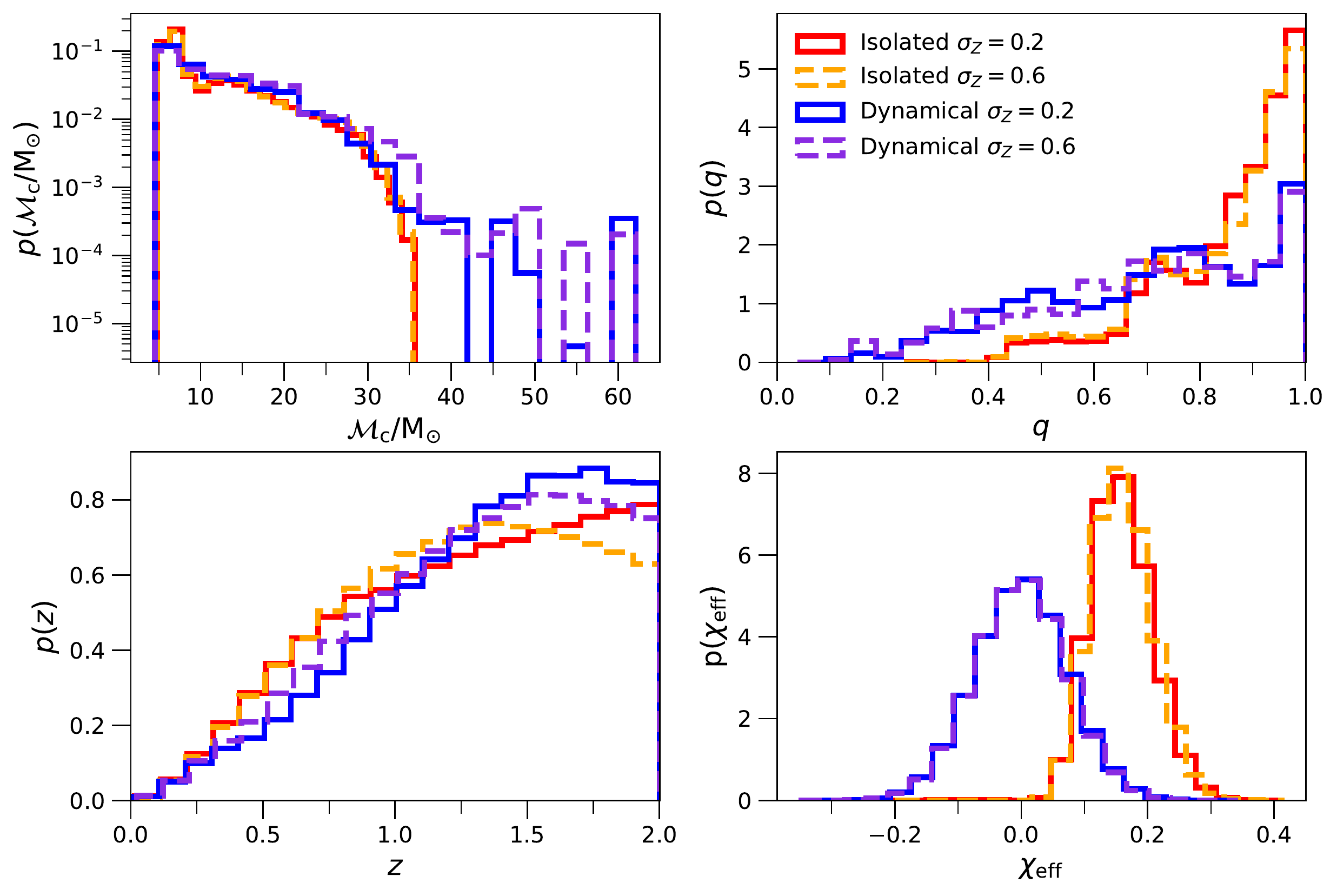}
\caption{Distribution of the model's parameters as inferred from a catalogue of $N_{\rm tot} = 3 \times 10^{5}$ sources. From left to right and from top to bottom, we plot the distribution of the chirp mass $\mathcal{M}_{\rm c}$, mass ratio $q$, redshift $z$ and effective spin  $\chi_{\text{eff}}$ for the isolated and dynamical formation channels. The spread in metallicity is equal to $\sigma_{\rm Z}=0.2$ ($\sigma_{\rm Z}=0.6$) for the solid (dashed) lines. The root-mean square of the Maxwellian for the spins is $\sigma_{\rm sp}=0.1$.}
\label{astro_model}
\end{figure*}

Figure \ref{astro_model} shows 
the sources in our catalogues, for both formation channels, assuming $\sigma_{\rm Z}= \lbrace 0.2, 0.6 \rbrace$ and $\sigma_{\rm sp}=0.1$.  Regarding the mass distribution, we observe features specific to each formation channel as it was already presented in previous work  \citep{dicarlo2020,rastello2020,santoliquido2020}. In particular, the dynamical formation channel allows for values of chirp mass as high as $\approx{}60$ M$_{\odot}$, while $\mathcal{M}_{\rm c}$ caps at $\approx{}35$ M$_{\odot}$ for the isolated model. Moreover, the distribution of mass ratios is more peaked towards $q\approx{}1$ in the isolated case, while dynamical interactions tend to produce more systems with lower mass ratios. Finally, the impact of $\sigma_{\rm Z}$ on masses' distribution is relatively small for both formation channels.

Both the metallicity spread $\sigma_{\rm Z}$ and the type of formation channel do have an impact on the shape of the redshift distribution. In particular, for a given formation channel, the peak of the redshift distribution is shifted towards lower values of $z$ as the value of $\sigma_{\rm Z}$ increases. This happens because the BBH merger rate strongly depends on progenitor's metallicity: BBH mergers are $\sim{}3-4$ ($\sim{}1-2$) orders of magnitude more efficient at low metallicity than at high metallicity for isolated (dynamical) binaries. Hence, a larger metallicity spread, which means a larger fraction of metal-poor stars at low redshift, implies more mergers at low redshift \citep{santoliquido2020b}. In addition, for a fixed value of $\sigma_{\rm Z}$, the distribution of redshift carries information on the formation channel. The peak of the distribution moves depending on the formation channel, with for instance a peak close to $z=1.3$ and $z=1.6$ for isolated and dynamical channels, respectively, at $\sigma_{\rm Z}=0.6$. Another important feature is represented by the different values of curvature and slope at low redshift ($z \in [0,1]$), which could already give us significant insights on the population of events observed by LIGO--Virgo.

Finally, 
we observe a significant difference in the distribution of $\chi_{\rm eff}$ depending on formation channels. 
In particular, the dynamical model has equal support for positive and negative values of $\chi_{\text{eff}}$, and has a spread close to $\sigma_{\text{sp}}$; the isolated model has a very strong support towards positive values and is centered around $0.15$. 
Our isolated model allows for sources with values of $\chi_{\text{eff}}$ as low as $-0.2$. However, only a handful of them are in the range $\chi_{\text{eff}} \leqslant 0$: $\lbrace 0.18\%,0.03\%, 0.007\%\rbrace$ for $\sigma_{\rm Z} = \lbrace 0.2, 0.4,0.6 \rbrace$, respectively. These sources do nevertheless have an importance from the analysis point of view, since they allow the isolated model to have some support for negative values of $\chi_{\text{eff}}$.  

\section{Bayesian inference}
\label{sec:Bayes}

Hierarchical Bayesian inference has proved to be an invaluable tool to estimate and constrain features of a population of merging compact objects. The approach we used has been described in previous studies \citep[e.g.,][]{mandel2018b,bouffanais2019}, so we only present here the main equations. 
Given an ensemble of $N_{\text{obs}}$ observations, $\lbrace h \rbrace^{k}$, the posterior distribution of the hyper-parameter $\lambda$ is described as an inhomogeneous Poisson distribution
\begin{eqnarray}
p(\lambda,N_{\lambda} | \lbrace h \rbrace^{k}) \sim \text{e}^{-\mu(\lambda)}\,{}  \pi(\lambda,N_{\lambda}) \prod_{k=1}^{N_{\text{obs}}}  N_{\lambda} \int_{\theta} \mathcal{L}^{k}(h^{k} | \theta) \,{} p(\theta | \lambda)\,{}{\rm d}\theta{},
\nonumber\\
\label{post_hier_model} 
\end{eqnarray}
where $\pi(\lambda,N_{\lambda})$ is the prior distribution on $\lambda$ and $N_{\lambda}$, $\mu(\lambda)$ is the predicted number of detections for the model and $\mathcal{L}^{k}(h^{k} | \theta)$ is the likelihood of the $k^{\text{th}}$ detection. The predicted number of detections is given by
\begin{equation}
\mu(\lambda) = N_{\lambda} \,{} \beta(\lambda),
\end{equation}
where $\beta(\lambda)$ is the detection efficiency of the model with hyper-parameter $\lambda$, that can be computed as
\begin{equation}
\beta(\lambda) = \int p(\theta | \lambda) \,{}p_{\text{det}}(\theta)\,{} \text{d} \theta,
\label{efficiency_det}
\end{equation}
where $p_{\text{det}}(\theta)$ is the probability of detecting a source with parameters $\theta$. This probability can be inferred by computing the optimal signal-to-noise ratio (SNR) of the source and comparing it to a detection threshold. In our case, we computed the optimal SNR using LIGO Livingston as a reference, for which we approximated the sensitivity using the averaged sensitivity over all detections for all three observing runs separately. Furthermore, by putting the detection threshold at $\rho_{\rm thr} = 8$, it was shown that this single-detector approximation is a good representation of more complex analysis with a network of detectors \citep{abadie2010,abbott2016_2,wysocki2018}.

The values for the event's log-likelihood were derived from the posterior and prior samples released by the LVC, such that the integral in eq.~\eqref{post_hier_model} is approximated with a Monte Carlo approach as
\begin{equation}
\int \mathcal{L}^{k}(h^{k} | \theta)\,{} p(\theta | \lambda)  \,{}\text{d} \theta \approx \dfrac{1}{N^{k}_{s}} \sum_{i=1}^{N^{k}_{s}} \dfrac{p(\theta^{k}_{i} | \lambda)}{\pi^{\phantom{ }k}(\theta^{k}_{i}) },
\label{approx_integral_likeli}
\end{equation}
where $\theta_{i}^{k}$ is the $i^{\text{th}}$ posterior sample for the $k^{\text{th}}$ detection and $N_{s}^{k}$ is the total number of posterior samples for the $k^{\text{th}}$ detection. To compute the prior term in the denominator, we also used Gaussian kernel density estimation.

Finally, we can also choose to neglect the information coming from the number of sources predicted by the model when estimating the posterior distribution. By doing so, we can have some insights on the impact of the rate on the analysis. In practice, this can be done by marginalising eq.~\eqref{post_hier_model} over $N_{\lambda}$ using a prior $\pi(N_{\lambda}) \sim 1 / N_{\lambda}$ \citep{fishbach2018}, which yields the following expression
\begin{eqnarray}
p(\lambda| \lbrace h \rbrace^{k}) \sim \pi(\lambda) \prod_{k=1}^{N_{obs}}  \left[ \dfrac{\int \mathcal{L}^{k}(d | \theta) \,{} p(\theta | \lambda)  \,{} \text{d} \theta}{\beta(\lambda)} \right],
\label{post_hier_model_marg} 
\end{eqnarray}
where the integral can be approximated in the same way as in eq.~\eqref{approx_integral_likeli} and $\beta(\lambda)$ is given by eq.~\eqref{efficiency_det}.

\section{Results}\label{sec:results}

Our primary goal is to put constraints on the value of the mixing fraction, $f \in [0,1]$, that controls the proportion of BBHs in our mixed model as

\begin{equation}
    p(\theta | f,\,{}\sigma_{\rm Z},\,{}\sigma_{\text{sp}}) = f\,{}  p(\theta | \text{iso},\,{}\sigma_{\rm Z},\,{}\sigma_{\text{sp}}) + (1-f)\,{} p(\theta | \text{dyn},\,{}\sigma_{\rm Z},\,{}\sigma_{\text{sp}}) 
    \label{mixed_model_eq}
\end{equation}
where $p(\theta | \text{iso},\sigma_{\rm Z},\sigma_{\text{sp}})$ and $p(\theta | \text{dyn},\sigma_{\rm Z},\sigma_{\text{sp}})$ are the distributions corresponding to the isolated and dynamical models, respectively. A value of $f=0$ ($f=1$) indicates that our mixed model is composed of BBHs formed only via the dynamical (isolated) channel.

We inferred only the distribution of the mixing fraction while keeping the other hyper-parameters constant. The analysis was then repeated for all combinations of $\left( \sigma_{\rm Z},\,{}\sigma_{\text{sp}} \right)$. To generate the posterior distribution for the mixing fraction, we have used a Metropolis-Hastings algorithm with a chain run for $10^{7}$ iterations. We discarded the first $10^{4}$ iterations (burn-in) and estimated the autocorrelation length of our chains, so that we could trim our original chains to obtain quasi-independent samples of the posterior distribution.

From a computational point of view, the form of eq.~\eqref{mixed_model_eq} allows us to decompose the number of events $N_{\lambda}$, the number of detections $\mu(\lambda)$ and the term in the integral of eq.~\eqref{post_hier_model}, as two distinct contributions from the isolated and dynamical models. Our strategy was to compute the values associated with the isolated and dynamical models a priori, for the three terms mentioned above, so that we were able to easily combine them for any values of $f$ when running the Monte Carlo Markov Chain (MCMC).

\subsection{Mixing fraction posterior distribution}

\begin{figure}
\includegraphics[width = 0.5\textwidth ]{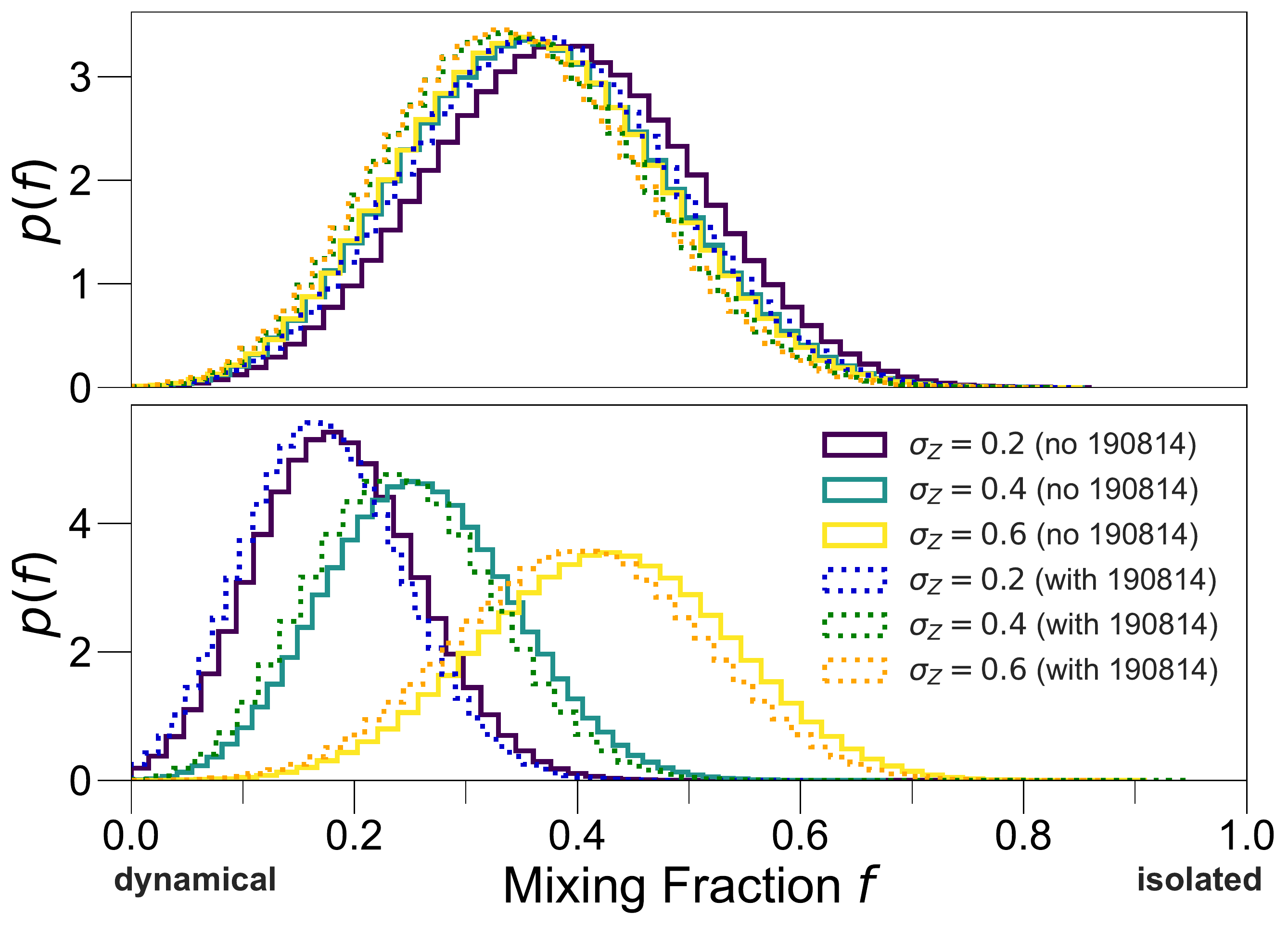}
\caption{Posterior distribution of the mixing fraction parameter as inferred from MCMC chains for $\sigma_{\text{sp}} =0.1$ and $\sigma_{\rm Z} = \lbrace 0.2,\,{}0.4,\,{}0.6 \rbrace$. The upper panel shows the distributions obtained when marginalising over the number of events (eq.~\eqref{post_hier_model_marg}), while the lower panel includes the rates (eq.~\eqref{post_hier_model}). The case in which we include (we do not include) GW190814 in the analysis is shown with a dotted (solid)  line.}
\label{results_mixingfrac_ns}
\end{figure}

Figure~\ref{results_mixingfrac_ns} shows the mixing fraction posterior distribution in the case $\sigma_{\text{sp}} = 0.1$. The distributions are inferred from trimmed MCMCs obtained using the marginalised posterior from eq.~\eqref{post_hier_model_marg} (top) and the posterior including the rates in eq.~\eqref{post_hier_model} (bottom). The Figure also shows two different analyses, depending if we did or not include the event GW190814 in the detection set. We will begin by discussing the case where the event is not included.

 \begin{table}
	\begin{center}
	\label{tab:models}
	\begin{tabular}{cccc} 
        \hline
        
		Formation channel & $\sigma_{\rm Z}$ & $N_{\text{det}}$\\
        \hline	
        Isolated & 0.2 & 2 \\
        Dynamical & 0.2 & 38 \\
        Isolated & 0.4 & 10 \\
        Dynamical & 0.4 & 48 \\
        Isolated & 0.6 & 45 \\
        Dynamical & 0.6 & 67 \\
        \hline
		
	\end{tabular}
	\caption{Number of detections as a function of formation channel and $\sigma_{\rm Z}$ for O1, O2 and O3a. We assumed $\sigma_{\text{sp}} =0.1$.}
	For reference, the actual number of BBH events during O1+O2+O3a is 44. We only consider the events presented in Table~1 of {\protect\cite{abbottO3apop}}.   

	\end{center}
	\label{table_ndet}
\end{table}

First, if we look at the case where rates are not included, we see that the three distributions corresponding to the three $\sigma_{\rm Z}$ are very similar and Gaussian-like. The medians of the distributions are equal to $0.39$, $0.36$ and $0.36$ for $\sigma_{\rm Z} = 0.2$, $0.4$ and $0.6$, respectively, indicating a slight preference towards the dynamical scenario. However, a pure dynamical scenario is outside the $99\%$ credible interval, for which we find lower bound values equal to $0.2$, $0.18$ and $0.17$ for increasing values of $\sigma_{\rm Z}$.

When taking into account the model rates, we do observe a significant change between the distributions, with medians located at $0.18$, $0.26$ and $0.43$ for $\sigma_{\rm Z}=0.2,$ 0.4 and 0.6. To better understand this behavior, Table~\ref{table_ndet} reports the values for the number of expected detections as a function of $\sigma_{\rm Z}$. The isolated model only predicts 2 detections at $\sigma_{\rm Z} = 0.2$, which is quite far away from the actual $44$ BBH mergers detected during O1, O2 and O3a. In comparison, the dynamical model has a better prediction of 38 detections, explaining why the posterior distribution of the mixing fraction shifts towards lower values when taking into account rates. In contrast, at $\sigma_{\rm Z} = 0.6$, the isolated model performs better than the dynamical one, with 38 predicted detections compared to 67. As a result, the posterior distribution is shifted towards higher values of the mixing fraction. 

Figure \ref{results_mixingfrac_ns} also shows the results obtained when including the event GW190814. This GW event is peculiar due to the low mass of its secondary component, with a median value of $m_{2}=2.59$~M$_{\odot}$, and is an outlier in the observed mass distribution. \cite{abbottO3apop} show that including the event in their analysis has a large impact on the mass distribution and rate inference. In fact, the median value of the BBH merger rate density rises from $23 \text{ Gpc}^{-3}\text{yr}^{-1}$ when ignoring GW190814, up to $58 \text{ Gpc}^{-3}\text{yr}^{-1}$ when the outlier is included. In our analysis, the inclusion of the event has only a little impact on the inference of the mixing fraction.  
This happens because our model distributions are fixed and do not depend on the set of detected events, unlike in \cite{abbottO3apop}, where model distributions do depend on considered detections. In the analysis presented by \cite{abbottO3apop}, an outlier  like GW190814 has a large impact on the inference of the mass distribution, which then impacts the detectable volume and the rate inference.

\begin{figure}
\includegraphics[width = 0.5\textwidth ]{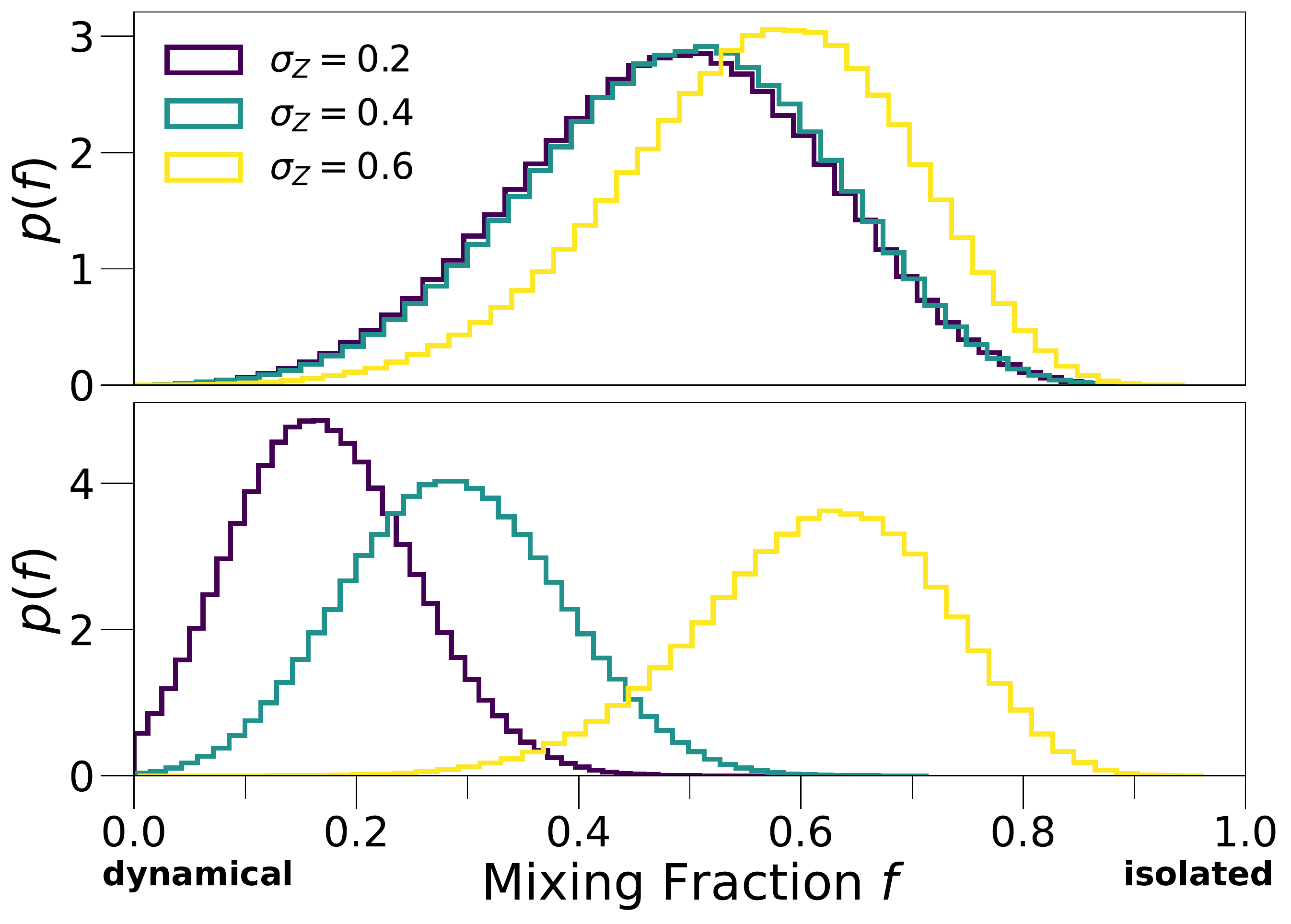}
\caption{Same as Figure~\ref{results_mixingfrac_ns}, but for $\sigma_{\text{sp}} =0.01$ (low-spin case). We show only the results we obtain not including GW190814 in our analysis.} 
\label{results_mixingfrac_ls}
\end{figure}

We now vary the parameter of the spin magnitude $\sigma_{\rm sp}$. Figure~\ref{results_mixingfrac_ls} shows the mixing fraction for $\sigma_{\text{sp}}=0.01$. Given the very low magnitude of the spin, most of the values for $\chi_{\text{eff}}$ are very close to 0 regardless of the formation channel, suggesting that the relevant parameters are restricted to the masses' parameters and redshift. In this case, the posterior distribution of the mixing fraction obtained marginalising over the rates shifts towards higher values. The resulting median values are equal to $0.49$, $0.49$ and $0.57$ for $\sigma_{\rm Z}=0.2$, 0.4 and 0.6. When taking the rates into account, we do observe the same pattern as for $\sigma_{\rm sp}$,  with a clear differentiation between the three $\sigma_{\rm Z}$ cases. This is dictated once more by the match between the expected number of detections and the actual number of detected events.

\begin{figure}
\includegraphics[width = 0.5\textwidth ]{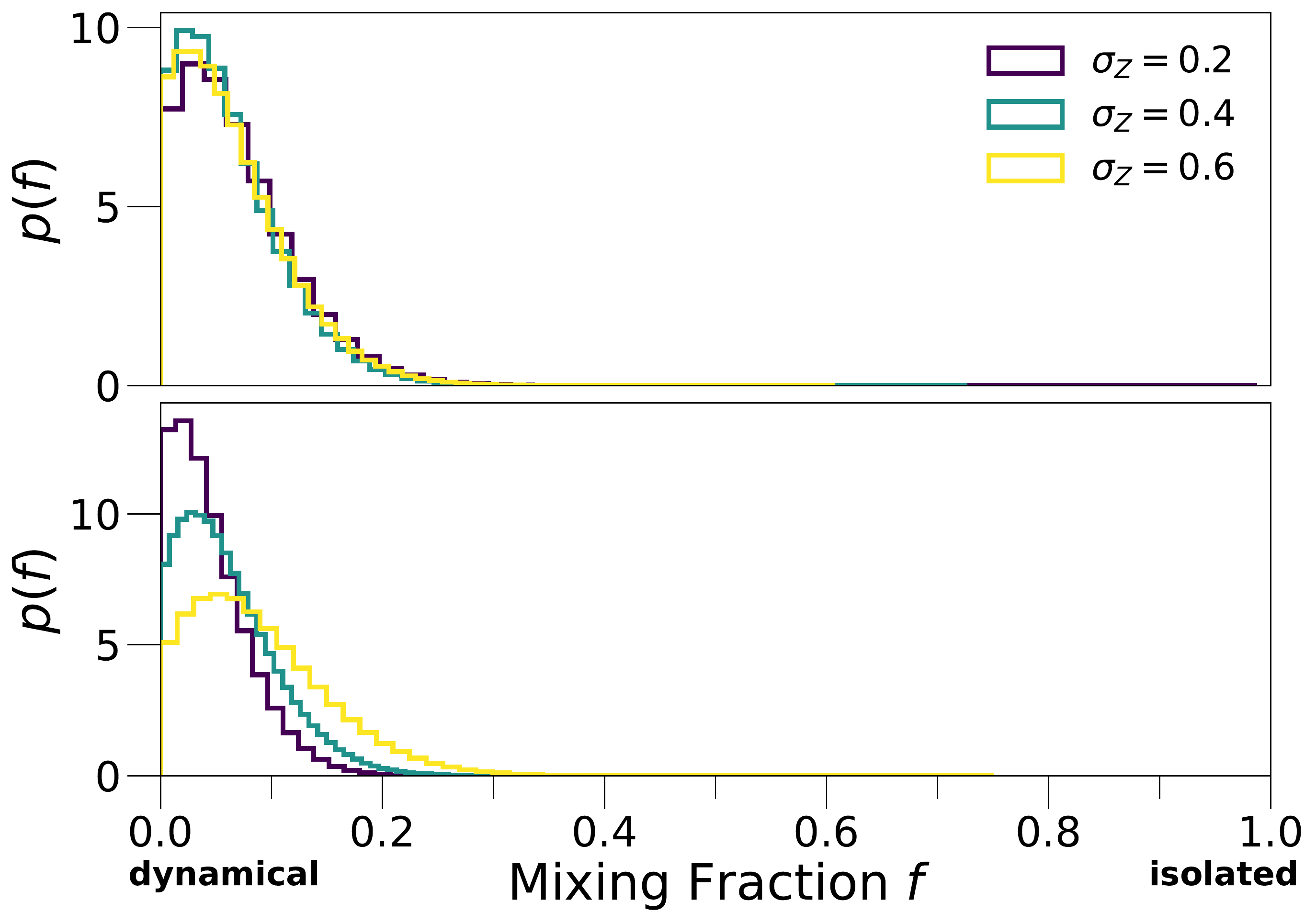}
\caption{Same as Figure~\ref{results_mixingfrac_ls}, but for $\sigma_{\text{sp}} =0.3$ (high-spin case).}
\label{results_mixingfrac_hs}
\end{figure}

Finally, Figure~\ref{results_mixingfrac_hs} shows the results for the high-spin case, in which $\sigma_{\rm sp}=0.3$. 
In this case, regardless of whether we include the rates or not in the analysis, all posterior distributions give strong support  towards the dynamical formation channel. This springs from the fact that the distribution of $\chi_{\text{eff}}$ in the isolated case peaks at positive values close to $0.3$, making it very difficult for this model to explain events with negative values of $\chi_{\text{eff}}$.

\begin{figure}
\includegraphics[width = 0.5\textwidth ]{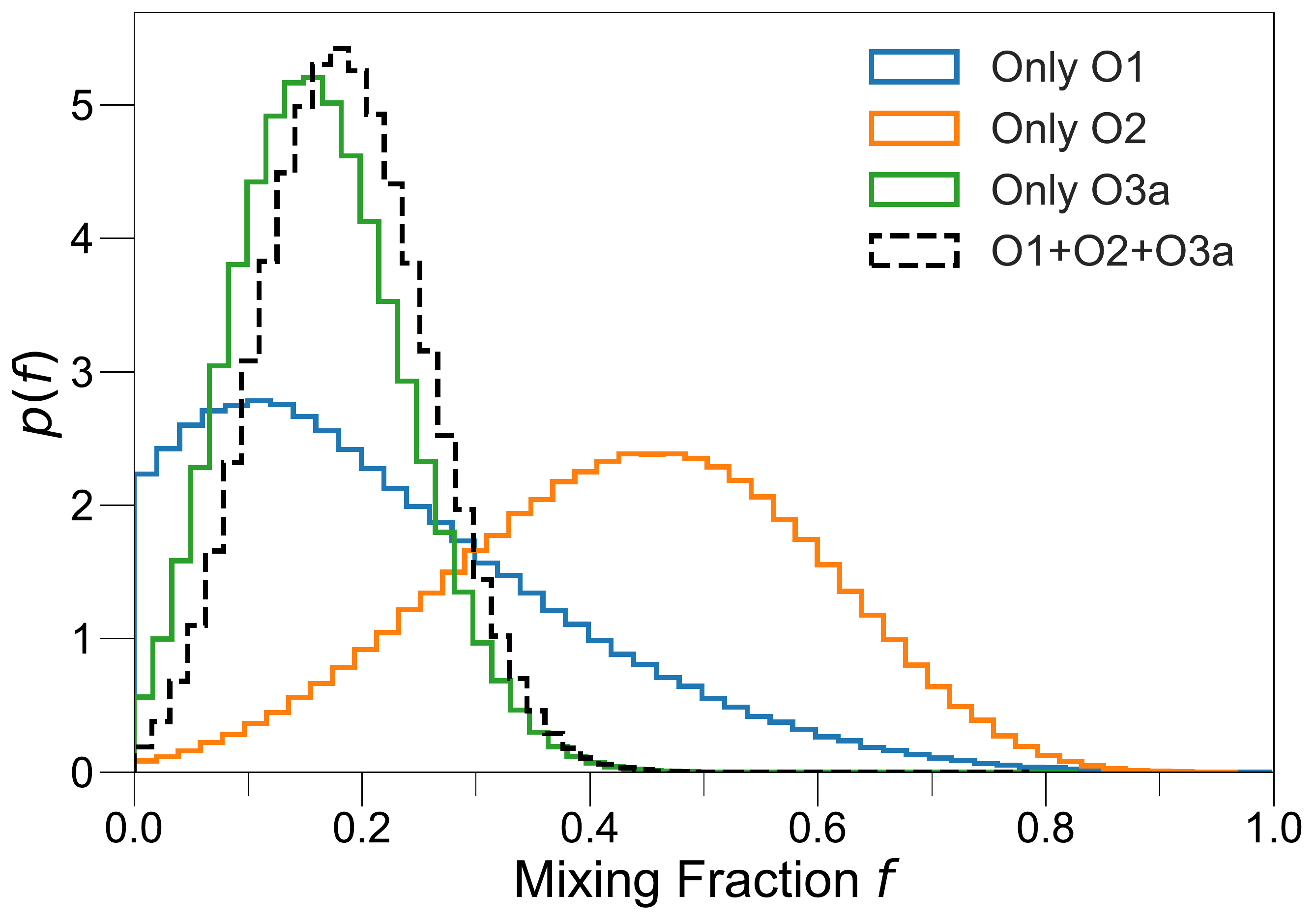}
\caption{Posterior distribution of the mixing fraction parameter when considering each of the observation runs separately. Blue line: O1; orange line: O2; green line: O3a. For reference, the dashed black line shows the case where all three runs are considered simultaneously. This result was obtained with $\sigma_{\rm Z}=0.2$ and $\sigma_{\text{sp}}=0.1$.}
\label{separate_run}
\end{figure}

In Figure \ref{separate_run}, we repeat the analysis for $\left( \sigma_{\text{sp}}=0.1, \sigma_{\rm Z} = 0.2 \right)$ but considering only the detected events in each of the observing runs separately. The width of the posterior distribution depends on  the number of events detected during the observing run, as expected. We also see that the median of the distribution went from $0.19$ for O1, to $0.44$ for O2 and $0.16$ for O3a. The important shift of the median is understandable given the relatively low number of events, especially for O1 and O2. For instance, the values of the integral from eq.~\eqref{approx_integral_likeli} that we found for GW150914 is an order of magnitude larger for the dynamical model compared to the isolated model. As only three GW events were detected during O1, this event alone is driving the posterior distribution towards a dynamically-dominated mixing model.

We can summarize our main results as follows.
\begin{itemize}
    \item For our fiducial spin distribution ($\sigma_{\rm sp}=0.1$), the median value of the mixing fraction is $f\sim{}0.35$ when we do not include the rates in our calculations, regardless of the metallicity spread $\sigma{}_{\rm Z}$. This shows that both isolated and dynamical BBHs are required to match GWTC-2, with a slight preference for dynamical BBHs. The main reason is that only dynamical BBHs have support for large values of $\mathcal{M}_{\rm c}$ and large negative values of $\chi_{\rm eff}$ (upper panel of Fig.~\ref{results_mixingfrac_ns}). 
    
    \item For the low-spin spin distribution ($\sigma_{\rm sp}=0.01$), the median value of the mixing fraction is $f\sim{}0.5-0.6$ when we do not include the rates in our calculations  (upper panel of Fig.~\ref{results_mixingfrac_ls}), indicating a stronger impact of the isolated channel than in the case of the fiducial spin distribution ($\sigma_{\rm sp}=0.1$). This shift to larger values of $f$ for smaller values of $\sigma_{\rm sp}$ is  an effect of the effective spin magnitude: very low values of $\sigma_{\rm sp}$ are associated with vanishingly small values of $\chi_{\rm eff}$.
    \item When we also account for the merger rate, the median value of the mixing fraction for both $\sigma_{\rm sp}=0.01$ (lower panel of Fig.~\ref{results_mixingfrac_ls}) and $\sigma_{\rm sp}=0.1$ (lower panel of Fig.~\ref{results_mixingfrac_ns}) strongly depends on the metallicity spread $\sigma_{\rm Z}$: lower values of the metallicity spread favour the dynamical channel over the isolated one, because the merger rate of isolated BBHs is too low if we assume the lowest metallicity spread ($\sigma_{\rm Z}=0.2$,Table~\ref{table_ndet}), while dynamical BBHs are less affected by the metallicity spread (see the discussion in \citealt{santoliquido2020}). Hence, the main impact of $\sigma_{\rm Z}$ is on the detection rates.
    \item In the large spin model ($\sigma_{\rm sp}=0.3$), the dynamical scenario is strongly favoured both if we include or if we neglect the rates (Figure~\ref{results_mixingfrac_hs}), because the values of $\chi_{\rm eff}$ for the isolated channel are too large to reconcile with GWTC-2 data.
\end{itemize}

\subsection{Best mixing model distribution}
 
\begin{figure*}
\includegraphics[width = 0.9\textwidth ]{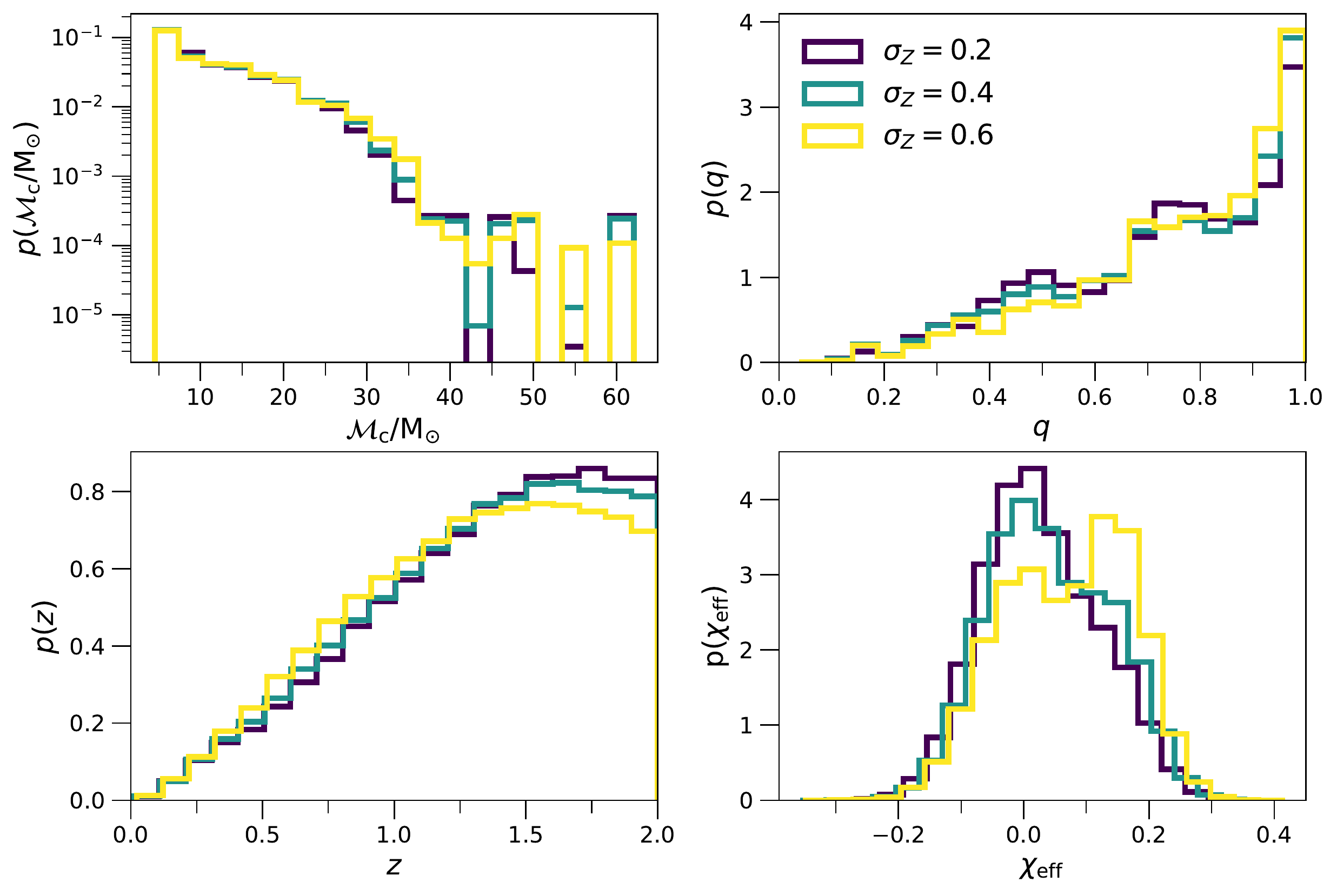}
\caption{Distribution of parameters for mixture models obtained using the median values of the mixing fraction inferred from the posterior distributions computed via our Bayesian analysis. We adopt  $\sigma_{\text{sp}}=0.1$, and we assume a mixing fraction equal to $f=0.18,$ 0.26 and $0.43$ for $\sigma_{\rm Z}=0.2,\,{}0.4$ and $0.6$, respectively. From left to right and from top to bottom, we plot the distribution of the chirp mass $\mathcal{M}_{\rm c}$, mass ratio $q$, redshift $z$ and effective spin  $\chi_{\text{eff}}$.}
\label{best_models}
\end{figure*}

 Figure \ref{best_models} shows the distribution of $\mathcal{M}_{\rm c}$, $q$ ,$z$ and $\chi_{\text{eff}}$ for our "best" mixing models for $\sigma_{\text{sp}}=0.1$, i.e. models where the mixing fraction is equal to the median of the posterior distribution derived in the previous analysis ($f=0.18$, 0.26 and 0.43 for $\sigma_{\rm Z}=0.2$, 0.4 and 0.6, respectively).
 
 For all values of $\sigma_{\rm Z}$, the chirp mass distribution has a tail at high masses ($\mathcal{M}_{\rm c} > 35$ M$_{\odot}$), that is populated by sources from the dynamical model. No substantial changes in the model distribution can be seen as a function of $\sigma_{\rm Z}$, especially for $\mathcal{M}_{\rm c} < 30$ M$_{\odot}$. We do observe some differences at higher values of $\mathcal{M}_{\rm c}$, but these are likely a statistical effect due to the limited size of our dynamical catalogues. As for chirp mass, the mass ratio distributions are very similar for all values of $\sigma_{\rm Z}$, and contain features from both the isolated (peak at $q$ close to 1) and the dynamical models (tail at low values of $q$).

In contrast, we do see a dependence of the distribution of $z$ as a function of $\sigma_{\rm Z}$, with a shift of the peak towards lower values of redshift for increasing values of $\sigma_{\rm Z}$. The reason  is that a higher percentage of metal-poor stars form at low redshift when the metallicity spread is larger, shifting the peak of merger redshifts to lower values.
 
 Finally, the distribution of $\chi_{\text{eff}}$ clearly shows the features of our fiducial spin model. The secondary peak at $\chi_{\text{eff}} \sim 0.15$ is more pronounced for higher values of $\sigma_{\rm Z}$, because the weight of the isolated formation channel is stronger at higher values of $\sigma_{\rm Z}$.
 
\begin{figure*}
\includegraphics[width = 0.9\textwidth ]{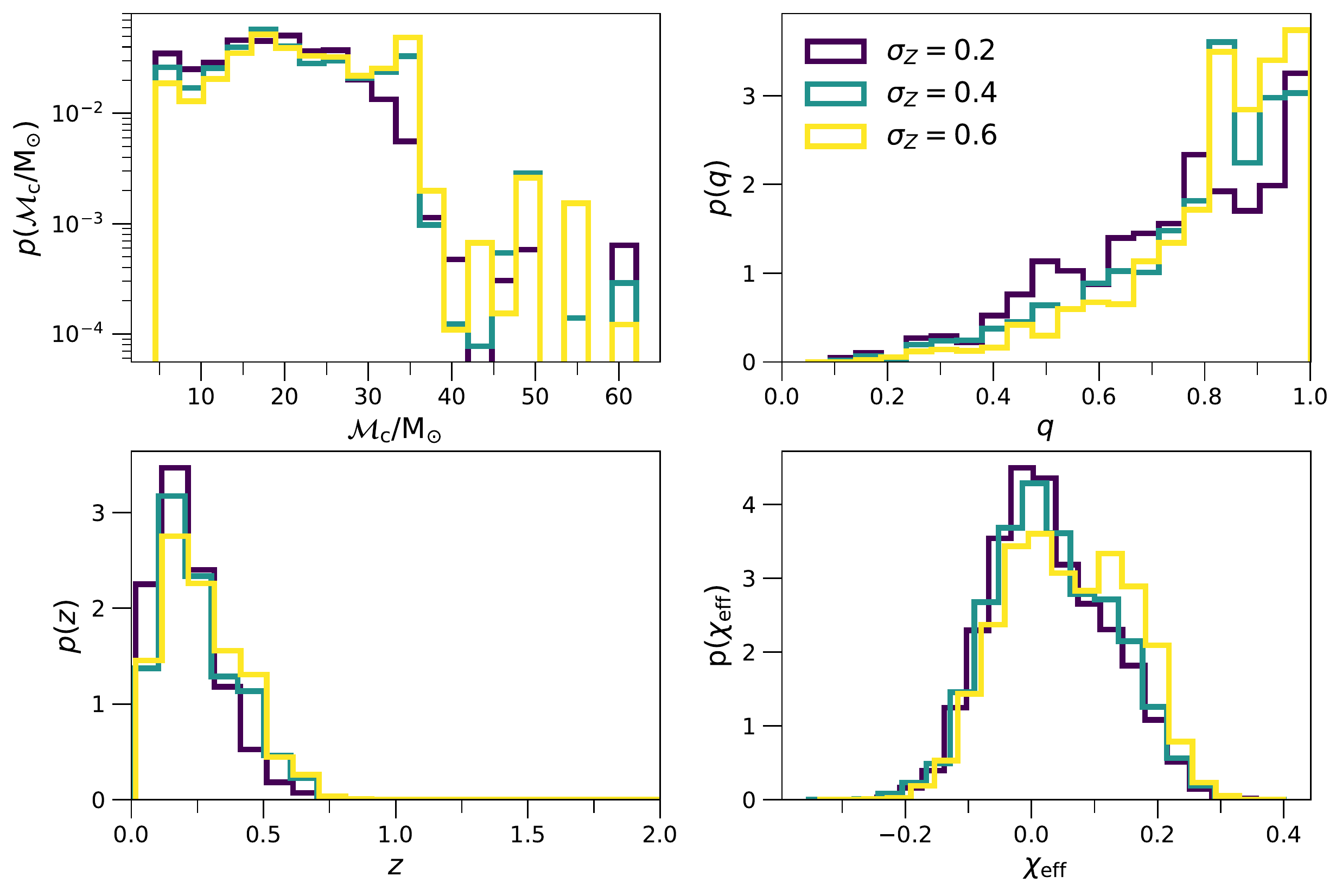}
\caption{Same as Figure \ref{best_models}, but weighted by the probability of detection $p_{\text{det}}$ considering the sensitivity of O3a.}
\label{best_models_detected}
\end{figure*}

 Figure \ref{best_models_detected} shows the same distributions as Figure \ref{best_models}, but  each source has been weighed by its probability of detection $p_{\text{det}}$ assuming the sensitivity of O3a. This selection effect can clearly be seen on the chirp mass distribution: the peak of the distribution is shifted from $10$ to $20$ M$_{\odot}$ regardless of the value of $\sigma_{\rm Z}$. In addition, we observe an increase in the number of very massive events with $\mathcal{M}_{\rm c} > 35$ M$_{\odot}$. We also see a very strong change in the distribution of redshift due to selection effects. The distributions now peak around $z\sim{0.3}$ and  do not have almost any support for $z>0.8$.
 
 Finally, we also observe some small changes in the distribution of $q$ and $\chi_{\text{eff}}$. In particular, we see a peak appearing around $q\sim{0.8}$ for $\sigma_{\rm Z}=0.4, 0.6$. This comes from the massive events in the dynamical model, that are the ones most easily detected. We see a diminution of the peak of $\chi_{\rm eff}$ at $0.15$ for $\sigma_{\rm Z}=0.6$. Once again, this happens because 
 our dynamical events are most easily observed than the isolated ones, due to their higher masses.

\section{Discussion}\label{sec:discussion}

We have considered a mixing model between the isolated formation channel and the dynamical formation channel in dense young star clusters. Our results exclude a purely isolated formation channel for the 44 BBH mergers in GWTC-2, at the 90\% credible level. According to our fiducial model ($\sigma_{\rm sp}=0.1$) and our low-spin case ($\sigma_{\rm sp}=0.01$), both the isolated channel and the dynamical one contribute to the events reported in GWTC-2. In contrast, the high-spin case ($\sigma{}_{\rm sp}=0.3$) strongly favours the dynamical channel with respect to the isolated one, because most spins are aligned with the orbital angular momentum in the latter model, while some events in GWTC-2 have support for negative values of $\chi_{\rm eff}$. In our analysis, we have considered just four parameters $\theta=\{\mathcal{M}_{\rm c},\,{}q,\,{}z,\,{}\chi_{\rm eff}\}$. We have neglected other observational parameters because including additional parameters would have made our analysis much slower from a computational perspective and because additional parameters are less informative based on current data. For example, the precessing spin parameter $\chi_{\rm p}$, which measures the main spin component in the orbital plane, is extremely important from a theoretical perspective, as large values of $\chi_{\rm p}$ are mainly associated with dynamical mergers and second-generation mergers \citep[e.g.,][]{mapelli2021}, but only a few events in GWTC-2 have significant constraints on $\chi_{\rm P}$ \citep[e.g.,][]{abbottGW190814}.

One of the key uncertainties of our analysis is the metallicity evolution of the stellar progenitors across cosmic time. We study its impact on our results by varying the parameter $\sigma_{\rm Z}$, corresponding to the metallicity spread. Larger values of $\sigma{}_{\rm Z}$ strengthen the contribution of the isolated channel (median value of $f\approx{}0.4$ in the fiducial case), while lower values of $\sigma_Z$ give more support to the dynamical scenario (median value of $f\approx{}0.2-0.3$). This trend is particularly strong when the merger rates are taken into account in our analysis, because the merger rate of the isolated channel is more sensitive to the metallicity of the progenitors than the dynamical one \cite[e.g,][]{santoliquido2020}. A better knowledge of the metallicity evolution across cosmic times is then crucial to infer the actual contribution of different formation channels.

Our results are in fair agreement with those of previous studies, comparing different BBH formation channels with GWTC-2. For example, \cite{wong2020} apply a Bayesian inference analysis to BBHs formed in the field 
and in globular clusters. 
Their analysis is done with the set of GWs' parameters $\left(m_{1}, m_{2}, z\right)$, marginalising over the number of events (i.e. computing the posterior distribution as in eq.~\eqref{post_hier_model_marg}) and adopting a constant value for the spread of metallicity, $\sigma_{\rm Z} = 0.5$. Similar to our conclusions, they find that the dynamical scenario is favoured, with a median value for the mixing fraction equal to $0.2$, but both formation channels are required to properly explain the observations. \cite{zevin2020} investigate an ensemble of formation channels that cover isolated 
and dynamical 
formation in globular clusters and nuclear star clusters. They marginalise over the number of events, use a fixed value for the metallicity spread ($\sigma_{\rm Z} = 0.5$) and  adopt the same parametrisation as we do:  $\theta{}=\left\{ \mathcal{M}_{\rm c},q,z,\chi_{\text{eff}} \right\}$. 
When they restrict their analysis to just  two channels (i.e. binaries evolved through common envelope and 
dynamical binaries formed in globular clusters), they find that the isolated scenario is favoured for most values of their spin-model parameters, except for the highest spin scenario. Their analysis also suggests that a combination of formation channels is  necessary to have the better representation of the observed events and that the higher spin case favours the dynamical scenario. 


To summarize, our main results and these of both \cite{wong2020} and \cite{zevin2020} point in the same direction: both isolated and dynamical scenarios are necessary to properly describe current observations from GWTC-2. 


\cite{wong2020} and \cite{zevin2020} focus on old massive globular clusters and nuclear star clusters, while we target dense young star clusters. This is the first time that the young star cluster scenario is compared against the new GWTC-2 data. Young star clusters are generally less massive ($\sim{}10^{2-5}$ M$_\odot$) and shorter lived ($\lesssim{}1$ Gyr) than both globular clusters and nuclear star clusters \cite[e.g.,][for a review]{portegieszwart2010}, but they are the most common birthplace of massive stars, especially in the local Universe \cite[e.g.,][]{lada2003}. Hence, their contribution to the local merger rate might be crucial, as already discussed by several authors \cite[e.g.,][]{banerjee2017,banerjee2020,dicarlo2020,kumamoto2020,santoliquido2020}. Dynamics in young star clusters acts in a different way with respect to globular clusters and nuclear star clusters. Firstly, young star clusters have much shorter two-body relaxation timescales than both globular and nuclear star clusters:
\begin{equation}
t_{\rm rlx}\sim{}10\,{}{\rm Myr}\,{}\left(\frac{M_{\rm SC}}{5000\,{}{\rm M}_\odot}\right)^{1/2}\,{}\left(\frac{r_{\rm vir}}{1\,{}{\rm pc}}\right)^{3/2},
\end{equation}
where $r_{\rm vir}$ is the virial radius. For young star clusters $t_{\rm rlx}$ is a few ten Myr, while it is several hundred Myr (or even a few Gyr) for the typical mass and size of globular clusters and nuclear star clusters. 
Hence,  the stellar progenitors of BHs have enough time to sink to the core of a young star cluster and to dynamically interact with each other, even before they collapse to BHs \citep{dicarlo2019a,kumamoto2019}. In contrast, massive stars die before they sink to the core in both globular clusters and nuclear star clusters. This explains why stellar collisions are more important in young star clusters than in other star clusters, possibly leading to the formation of BHs with masses $>65$ M$_\odot$, such as the ones we considered here \citep{portegieszwart2004,dicarlo2019b}.

Secondly, the majority of massive binary stars are hard in young star clusters (i.e., they have a binding energy higher than the average kinetic energy of a star in the cluster, \citealt{heggie1975}). In contrast, most binary stars are soft in globular clusters and nuclear star clusters. This has a crucial impact on the formation channel of BBHs: most original binary stars break in globular/nuclear star clusters because of dynamical encounters, and most BBHs  form from interactions among three single BHs \cite[three-body captures, e.g.,][]{morscher2015,antonini2016}. In contrast, most binary stars survive in young star clusters: they harden by flybys and undergo dynamical exchanges. Hence, most BBHs in young star clusters form by dynamical exchanges rather than three-body captures. 

Thirdly, the escape velocity from young star clusters is $\sim{}10$ km s$^{-1}$, significantly smaller than the one of globular ($\sim{}30$ km s$^{-1}$) and nuclear star clusters ($\sim{}100$ km s$^{-1}$, \citealt{antonini2016}). Hence,  hierarchical mergers among BBHs are extremely rare in young star clusters, while they are common in nuclear star clusters \citep{mapelli2021}. For all of these differences among young, globular and nuclear star clusters, the properties of BBHs in young star clusters are quite peculiar and deserve more investigation. In a follow-up study, we will compare the populations of BBHs in young star clusters, globular clusters and nuclear star clusters together, in order to remove the bias of having only two formation channels \citep{zevin2020}.

Finally, the mass function of BBHs is certainly one of the key ingredients of our results. In young star clusters, we have merging BBHs with primary masses up to $\sim{90}$ M$_\odot$, while the maximum primary mass in isolated BBH mergers is $\sim{}40$ M$_\odot$. This difference is primarily connected with the collapse of the hydrogen envelope of the progenitor star. In tight stellar binaries, non-conservative mass transfer and common envelope lead to the complete ejection of the hydrogen envelope, before the formation of the second BH. Hence, the maximum BH mass in merging isolated binaries is $\sim{}40$ M$_\odot$, corresponding to the maximum helium core mass below the pair instability threshold. In contrast, massive single stars and massive stars in loose binary systems (orbital separation $\gtrsim{}10^4$ R$_\odot$) can preserve a fraction of their initial hydrogen envelope to the very end, leading to the formation of BHs with masses up to $\sim{}65$ M$_\odot$ \citep[e.g.,][]{mapelli2020,costa2021}. In isolation, these BHs do not merge, but in star clusters they can pair up dynamically and lead to massive BBH mergers. In addition, dynamically triggered collisions between massive stars can lead to even more massive BHs, up to $\sim{}90$ M$_\odot$. While primary BHs with mass $>60$ M$_\odot$ represent only $\sim{}0.5$\% of all BBH mergers in our simulations \citep{dicarlo2019b}, they give a crucial contribution to the current analysis.

\section{Summary}\label{sec:summary}

The LVC has recently published the second GW transient catalogue (GWTC-2, \citealt{abbottO3a}), increasing the number of binary compact object mergers from 11 to 50 events, most of them BBH mergers. This large number of events makes it possible to obtain the first constraints on the formation channels of BBHs.

Here, we explore two alternative formation channels: i) isolated binary evolution via stable mass transfer and common envelope, ii) dynamical formation in dense young star clusters. Young star clusters are generally less massive and shorter lived than globular clusters, but they are the most common birthplace of massive stars \citep[see][for a review]{portegieszwart2010}. Hence, most BHs are likely born  in dense young star clusters.

Comparing our simulations to GWTC-2, we estimate the mixing fraction, i.e. the fraction of BBHs formed in young star clusters versus isolated binaries. Assuming that the spin magnitudes follow a Maxwellian distribution, we consider three different models for the spin, corresponding to a standard deviation parameter $\sigma_{\rm sp}=0.01,$ 0.1 and 0.3 in the low, fiducial and high spin case. Finally, we probe the impact of metallicity evolution, by varying the metallicity spread parameter $\sigma_{\rm Z}=0.2,$ 0.4, 0.6.

We find that the isolated binary evolution scenario struggles to match all the events listed in GWTC-2. In the low-spin and fiducial spin models, a mixture of both isolated and dynamical binaries is needed to account for GWTC-2 events. Finally, the high-spin case has a strong preference for the dynamical channel, mostly because of the support for negative values of $\chi_{\rm eff}$ in some GWTC-2 events. 

The metallicity spread $\sigma_{\rm Z}$ is a key ingredient. For a fixed mean value of the stellar metallicity distribution, a large (small) metallicity spread tends to favour the isolated (dynamical) channel versus the dynamical (isolated) scenario. This confirms that more observational constraints on the evolution of stellar metallicities across cosmic time are urgently needed, to narrow down the uncertainties on BBH merger rates. Despite the large uncertainties on spin magnitudes and metallicity spread, our results point towards an exciting direction: more than one formation channel is needed to explain the properties of BBHs in the second GW transient catalogue, and the dynamical path is essential to account for the largest chirp masses and for negative values of the effective spin.

%
\section*{Acknowledgements}
We thank the anonymous referee and Simone Bavera for their useful comments, which helped us improving the manuscript. MM, YB, FS, UND, NG, SR and GI acknowledge financial support from the European Research Council for the ERC Consolidator grant DEMOBLACK, under contract no. 770017. MCA and MM acknowledge financial support from the Austrian National Science Foundation through FWF stand-alone grant P31154-N27.  NG acknowledges financial support from the Leverhulme Trust Grant No. RPG-2019-350 and Royal Society Grant No. RGS-R2-202004.

\section*{Data availability}
The data underlying this article will be shared on reasonable request to the corresponding authors.

\bibliographystyle{mnras}

\bibliography{DynVSIso_2020}

\appendix{}
\section{Changing the mean value of the metallicity distribution}\label{sec:appendix}
Here, we consider an alternative definition of the probability distribution of metallicity:
\begin{equation}
\label{eq:pdf2}
p(z', Z) = \frac{1}{\sqrt{2 \pi\,{}\sigma_{\rm Z}^2}}\,{} \exp\left\{{-\,{} \frac{\left[\log{(Z(z')/{\rm Z}_\odot)} - \langle{}\log{Z(z')/Z_\odot}\rangle{}\right]^2}{2\,{}\sigma_{\rm Z}^2}}\right\},
\end{equation}
where
\begin{equation}
    \langle{}\log{Z(z')/Z_\odot}\rangle{}=\log{\langle{}Z(z')/Z_\odot\rangle{}}-\frac{{\ln(10)}\,{}\sigma_{\rm Z}^2}{2}.
    \label{eq:average_Z}
\end{equation}
The above equation is formally correct \citep{bavera2020a}, given the definition of the mass-weighted metallicity in \cite{madau2017}, but implies that the mean of the log-normal distribution changes with the assumed value of $\sigma_{\rm Z}$. Namely, $\langle{}\log{Z(z=0)/Z_\odot}\rangle{}=0.107,$ $-0.031$ and $-0.261$ if $\sigma_{\rm Z}=0.2,$ 0.4 and 0.6, respectively. In the main text, we decided to keep the mean value fixed for the sake of simplicity, while here we discuss what happens if the mean value changes, too.

\begin{figure}
\includegraphics[width = 0.5\textwidth ]{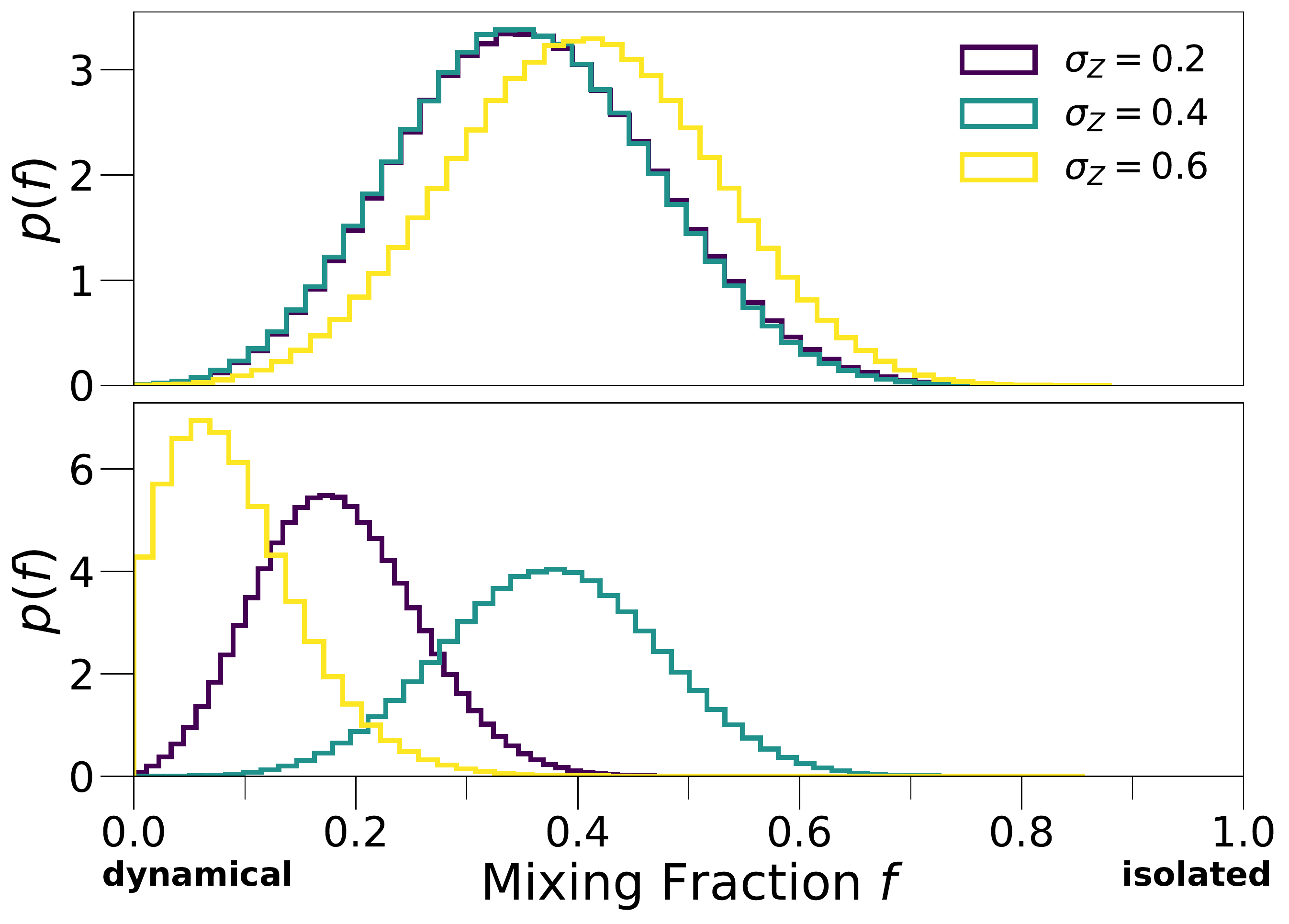}
\caption{Posterior distribution of the mixing fraction parameter as inferred from MCMC chains for $\sigma_{\text{sp}} =0.1$ and $\sigma_{\rm Z} = \lbrace 0.2,\,{}0.4,\,{}0.6 \rbrace$ using the alternate definition of metallicity distribution from eq. \eqref{eq:pdf2}}
\label{results_mixingfrac_ns_modMet}
\end{figure}

\begin{figure}
\includegraphics[width = 0.5\textwidth ]{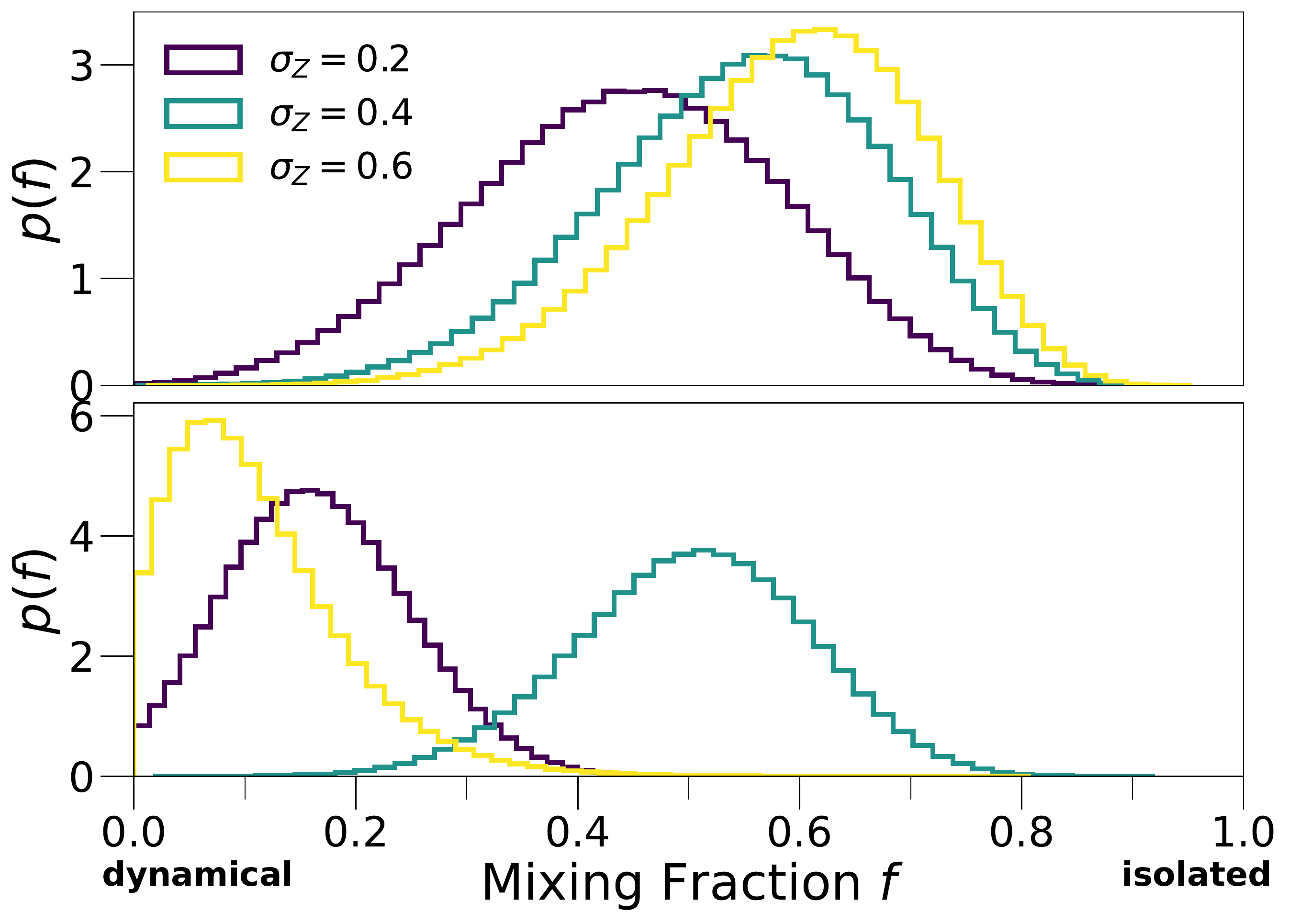}
\caption{Same as Figure~\ref{results_mixingfrac_ns_modMet}, but for $\sigma_{\text{sp}} =0.01$ (low-spin case).}
\label{results_mixingfrac_ls_modMet}
\end{figure}

\begin{figure}
\includegraphics[width = 0.5\textwidth ]{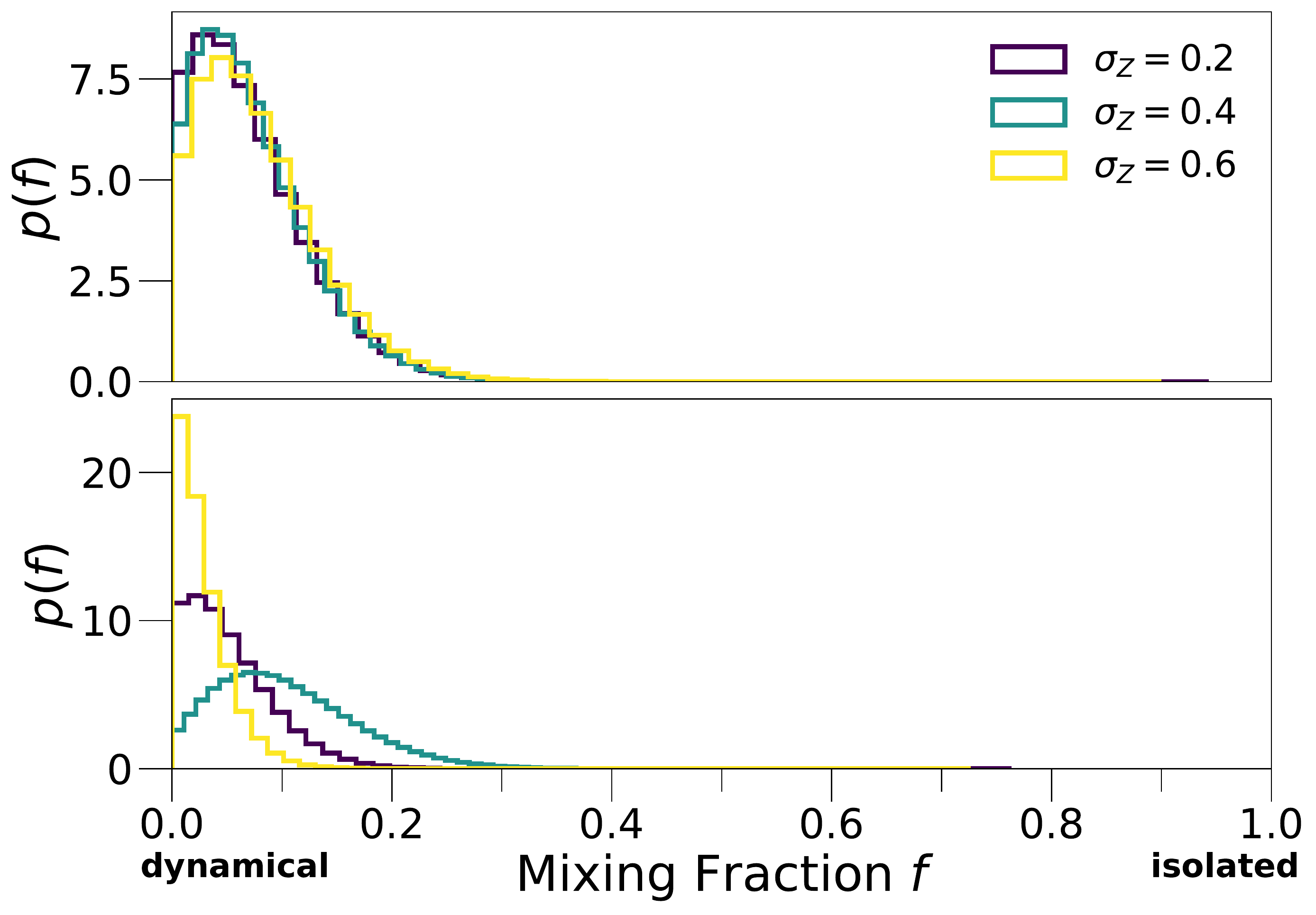}
\caption{Same as Figure~\ref{results_mixingfrac_ns_modMet}, but for $\sigma_{\text{sp}} =0.3$ (high-spin case).}
\label{results_mixingfrac_hs_modMet}
\end{figure}

Figures~\ref{results_mixingfrac_ns_modMet}, \ref{results_mixingfrac_ls_modMet} and \ref{results_mixingfrac_hs_modMet} show the mixing fraction we obtain by adopting this metallicity definition in the fiducial, low-spin and high-spin cases. Almost nothing changes for the high-spin case: the dynamical scenario is still highly favoured, because the data strongly disfavour high and aligned spins. In the low spin and fiducial spin models, we observe an important difference: the case with a large metallicity spread ($\sigma_{\rm Z}=0.6$) strongly favours the dynamical channel when the rates are included in our analysis (bottom panels of Figures~\ref{results_mixingfrac_ns_modMet} and \ref{results_mixingfrac_ls_modMet}). In contrast, in Figures~\ref{results_mixingfrac_ns} and \ref{results_mixingfrac_ls}, the case with a large metallicity spread has a preference for the isolated channel. The main reason for this difference is the strong dependence of the BBH merger rate on stellar metallicity in the isolated channel. When we assume both a large value of $\sigma_{\rm Z}=0.6$ and a low value of the mean of the metallicity distribution ($\langle{}\log{Z(z=0)/Z_\odot}\rangle{}=-0.261$, eq.~\eqref{eq:average_Z}), the number of expected detections in the isolated channel becomes very high (Table~\ref{table_ndet_modMet}). Such a large number of events is in tension with the observations. In contrast the rate of the dynamical channel is less affected by progenitor's metallicity \citep{santoliquido2020}. Hence, the dynamical channel ends up being more favoured if we assume a low mean value and a large spread of the metallicity distribution. This is a further confirmation that our results are extremely sensitive to the metallicity distribution.

\begin{table}
	\begin{center}
	\begin{tabular}{cccc} 
        \hline
        
		Formation channel & $\sigma_{\rm Z}$ & $N_{\text{det}}$\\
        \hline	
        Isolated & 0.2 & 3 \\
        Dynamical & 0.2 & 38 \\
        Isolated & 0.4 & 25 \\
        Dynamical & 0.4 & 64 \\
        Isolated & 0.6 & 169 \\
        Dynamical & 0.6 & 135 \\
        \hline
		
	\end{tabular}
	\caption{Number of detections as a function of formation channel and $\sigma_{\rm Z}$ for O1, O2 and O3a and $\sigma_{\text{sp}} =0.1$, using the alternate definition of metallicity distribution from eq. \eqref{eq:pdf2}. 
	For reference, the actual number of BBH events during O1+O2+O3a is 44. We only consider the events presented in Table~1 of {\protect\cite{abbottO3apop}}.  } 

	\end{center}
	\label{table_ndet_modMet}
\end{table}

\end{document}